\documentclass[aps,prd,epsf,showpacs,amsmath,amssymb, 11pt]{revtex4}
\usepackage{dcolumn}
\usepackage{bm}

\newcommand{\e}{equation$\;$} 
\newcommand{\M}{\ensuremath{{\cal M}(r,v)}}
 
\newcommand{\be}{\begin{equation}}
\newcommand{\ee}{\end{equation}}
\newcommand{\ba}{\begin{eqnarray}}
\newcommand{\ea}{\end{eqnarray}}
\newcommand{\ban}{\begin{eqnarray*}}
\newcommand{\ean}{\end{eqnarray*}}

\newcommand{\n}[1]{\label{#1}}

\newcommand{\eq}[1]{(\ref{#1})}

\newcommand{\mo}{\ensuremath{m_0}}
\newcommand{\mt}{\ensuremath{{\cal M}_2(v)}}
\newcommand{\mtv}{\ensuremath{{\cal M}_2(v)_{,v}}}

\newcommand{\X}{\ensuremath{{\cal X}}}

\newcommand{\ntr}{\ensuremath{{\nu(t,r)}}}
\newcommand{\str}{\ensuremath{\psi(t,r)}}
\newcommand{\ph}{\ensuremath{{\phi}}}
\newcommand{\dw}{\ensuremath{{d\Omega^2}}}

\newcommand{\f}{\ensuremath{f(r)}}
\newcommand{\E}{{\mathbb E}}
\newcommand{\G}{\ensuremath{{\cal G}}}

\begin{document}

\title{Gravitational collapse of a minimally coupled massless scalar field}

\author{Swastik Bhattacharya$^\ast$, Rituparno Goswami$^\dagger$, Pankaj S. 
Joshi$^\ast$}
\affiliation{$^\ast$ Tata Institute for Fundamental Research, Colaba, 
Mumbai 400005, India}
\affiliation{$^\dagger$ Department of Mathematics and Applied Mathematics, 
University of Cape Town, Cape Town, South Africa}

\begin{abstract} We study here the evolution of a massless
scalar field in a spacetime, developing from a regular initial spacelike 
surface. The Einstein equations and regularity and boundary conditions 
governing the same are specified. Both homogeneous and inhomogeneous 
collapse models are considered and we analyze when the occurrence of 
singularity will be simultaneous or otherwise. In the inhomogeneous 
collapse case, we characterize a wide family of black hole solutions 
arising in scalar field collapse. We also discuss the possibility of existence 
of classes of non-singular models where collapse could almost freeze
if suitable conditions are satisfied.

\end{abstract}
\maketitle

\section{Introduction}

The formation of spacetime singularities in gravitational
collapse and formation of black holes is an issue of great importance 
in gravitation physics, which has been investigated in much detail in 
Einstein's theory. The occurrence of singularities offers the regime where 
gravity is extreme, and where the quantum gravity effects would be important. 
As is known, dynamical evolution of matter fields in a spacetime generically 
yields a 
singularity, provided reasonable physical conditions are satisfied such as 
the causality, a suitable energy condition ensuring the positivity of energy 
density, and formation of trapped surfaces.

The case of gravitational collapse of a massless scalar field 
is of particular interest in both collapse situations as well as cosmological
scenarios. In cosmology, special importance is attached to the 
evolution of a scalar field, which has attracted a great deal of attention 
in past decades. This is because one would like to know the behaviour 
for fundamental matter fields towards understanding the transition from
matter dominated regime to dark energy domination (see e.g. 
~\cite{nunes} 
and references therein). Scalar fields are of much interest in 
view of the inflationary scenarios that govern the early universe dynamics 
because such a field can act as an `effective' cosmological constant in 
driving the inflation 
~\cite{nde}.
In gravitational collapse studies, the nature of 
singularity for massless scalar fields has been examined and a number
of numerical and analytical works have been done in recent years on 
spherical collapse models 
~\cite{scalar1}-\cite{scalar8},
from the perspective of the cosmic censorship hypothesis. The massless free 
scalar field has been studied for the static case also in some detail
( see e.g. \cite{JNW},
\cite{Wyman}, \cite{Chase}, \cite{Xan}, \cite{Bergman}, \cite{Buchdal}).
However, the dynamical case is more important to understand because 
that may give insights into phenomena such as gravitational collapse 
and the cosmic censorship hypothesis, other than the early universe 
and cosmological considerations.

In the present study, we first develop here a mathematical structure 
in a general manner to deal with the evolution of massless scalar 
fields in a spacetime. This would be applicable to problems either 
in cosmology or for gravitational collapse final states in a spherically 
symmetric spacetime.  We then examine the classes of collapsing 
models where the singularity in future forms simultaneously as the 
collapse develops, even when the density of the field could be 
inhomogeneous. We study here both homogeneous as well as 
inhomogeneous collapse models. We then characterize here  
a wide class of black hole models forming as collapse final state.  Apart 
from the spherical symmetry, 
we do not assume here any further constraints on the spacetime, such as the 
presence of a homothetic Killing vector, homogeneous or shearfree 
nature for the fluid, or such other conditions.

We note here the classes of models where our considerations would apply.
The analysis in this paper has been done using a comoving coordinate system.
For massless scalar fields, such a coordinate system would break down when 
the gradient of the scalar field becomes null. So the conclusions based on the 
present analysis would be valid for those classes of models and solutions 
for which the gradient of the scalar field is timelike throughout the dynamical evolution
of the field.  We note that the homogeneous and isotropic Friedman-Roberson-Walker 
solutions with a massless scalar field as the matter content is such an example,
where this condition is satisfied. Massless scalar field solutions with inhomogeneous 
perturbations around a homogeneous background would also be in this class. 
In fact, as we shall argue later, this class where the gradient of the scalar field
always remains timelike, would include a large number of physically relevant situations.
On the other hand, our analysis does not apply to the case where
the gradient of the scalar field changes its sign, becoming null and then
spacelike, from its original nature of being timelike. However, as we have 
argued here, the present class is of sufficient physical interest in its own right
to carry out an analysis of the same as far as the dynamical evolutions of
the scalar fields are concerned. A related important point that we note is, 
the analysis given here would also hold and describe the evolution of
stiff fluids in a spacetime in a general manner. This is because,
a massless scalar field with a timelike gradient, which is  minimally 
coupled to gravity, has an exact correspondence with a stiff 
fluid minimally coupled to gravity.

The massless scalar field $\ph(x^a)$ on a spacetime  
$(M, g_{ab})$ is described by the Lagrangian, 
\be
{\cal L}=-\frac{1}{2}\ph_{;a}\ph_{;b}g^{ab}.
\n{lag}
\ee
The corresponding Euler-Lagrange equation is 
$\ph_{;ab}g^{ab}=0$,
and the energy-momentum tensor for the scalar field, as calculated 
from the above Lagrangian is, 
\be
T_{ab}=\ph_{;a}\ph_{;b}-\frac{1}{2}g_{ab}\left(\ph_{;c}\ph_{;d}
g^{cd}\right).
\n{emt}
\ee
A massless scalar field is a {\it Type I} matter field
~\cite{haw},  
{\it i.e.}, it admits one timelike and three spacelike eigen vectors.
At each point $q\in M$, we can then express the tensor $T^{ab}$ 
in terms of an orthonormal basis $(\E_0,\E_1,\E_2,\E_3)$, where $\E_0$ 
is a timelike eigenvector with an eigenvalue $\rho$ and $\E_{\alpha}$ 
$(\alpha=1,2,3)$ are three spacelike eigenvectors with eigenvalues 
$p_\alpha$. The eigenvalue $\rho$ represents the energy density
of the scalar field as measured by an observer whose world line 
at $q$ has an unit tangent vector $\E_0$ and the eigenvalues $p_\alpha$ 
represent the principal pressure in three spacelike directions 
$\E_\alpha$.
We now choose the spherically symmetric coordinates $(t,r,\theta,\phi)$ 
along the eigenvectors $(\E_0,\E_\alpha)$, such that the reference frame 
is {\it comoving}. As discussed in 
~\cite{landau}, 
the general spherically symmetric metric in comoving coordinates
can be written as,
\begin{equation}
ds^2=-e^{2\ntr}dt^2+e^{2\str}dr^2+R^2(t,r)\dw,
\label{metric}
\end{equation}
where $\dw$ is the metric on a unit 2-sphere and 
we have used the two gauge freedoms of two variables, namely, 
$t'=f(t,r)$ and $r'=g(t,r)$, to make the $g_{tr}$ term in metric
and the radial velocity of the matter field to vanish. We note that we 
still have two scaling freedoms of one variable in $t$ and $r$.

In general, we have $\ph=\ph(t,r)$, but
from \e \eq{emt} it is easily seen that in the comoving reference frame 
\eq{metric}, we must have $\ph(t,r)=\ph(t)$ or $\phi(t,r)=\phi(r)$, 
because the energy-momentum tensor is diagonal. As we would like 
to investigate here the dynamic behaviour of the scalar field, 
we consider the former option. In this comoving frame the components 
of the energy-momentum tensor are,
\begin{equation}
T^t_t=T^r_r=T^{\theta}_{\theta}=T^{\phi}_{\phi}=
\frac{1}{2}e^{-2\ntr}\dot{\ph}^2\;\;.
\label{eq:em}
\end{equation}
Thus, we see that in the comoving frame the massless scalar 
field behaves like 
a {\it stiff} isentropic perfect fluid with the equation of state 
\be
p(t,r)=\rho(t,r)= e^{-2\ntr}\dot{\ph}^2/2.
\n{rhop}
\ee
We can easily see that for any real valued function 
$\ph(t)$, all energy conditions are satisfied by the matter field. 
We note that if we consider the field $\phi=\phi(t)$, then 
the weak energy condition guarantees that $\phi,_\mu $ is either timelike 
or null always in general. However, in a comoving frame, the gradient 
of the scalar field remains timelike throughout the collapse,
which is a property of the comoving frame by definition.

In a  physically reasonable collapse situation, the energy 
density of the matter field and the magnitude of the Ricci scalar are  
expected to increase with time. For a massless scalar field
we have $\mid \phi,_\mu \phi^{,\mu}\mid = 2\rho$. 
In such cases, if collapse progresses from a regular spacelike hypersurface 
where the gradient of the scalar field is timelike, then the density is 
non-zero initially and it can only increase afterwards. This would mean 
that throughout the collapse evolution of the scalar field from the given initial 
regular surface, the gradient of the scalar field would always remain 
timelike. So in this case of a collapsing scalar field, we can use the 
comoving coordinate system without any loss of generality and 
without a concern on a possible breakdown of the 
coordinate system. 

It is important to note that we are not specifying 
here the class of initial data from which one can have collapse situations 
where the density would always increase. But given such a solution, our 
analysis would hold. Therefore, the comoving system can be used to study 
the collapse of a massless scalar field, where the density either 
increases or does not decrease with time. In this paper, by a collapsing 
scalar field, we would mean only such cases.

The correspondence here with a stiff fluid can be seen from \eqref{emt}.
The energy momentum tensor for the perfect fluid is 
\begin{equation}
 T_{ab}= (\rho+p)u_a u_b + p g_{ab} \label{pf},
\end{equation}
where $u^\mu$ is the velocity vector.
For stiff fluid, $p= \rho$.\\
Since $\phi,_\mu$ is timelike, $\phi,_\mu \phi^{,\mu}= - \mid \phi,_\mu \phi^{,\mu} \mid $.
Defining $u_\mu=\frac{\phi,_\mu}{\mid \phi,_\mu \phi^{,\mu} \mid^{\frac{1}{2}}}$, the 
energy momentum tensor for the massless scalar field can be expressed as,
 $T_{ab}= (\mid \phi,_\mu \phi^{,\mu} \mid) u_au_b + \frac{1}{2} g_{ab} (\mid \phi,_\mu \phi^{,\mu} \mid)$.

Denoting $\mid \phi,_\mu \phi^{,\mu} \mid=\rho=p$, this expression for the energy momentum 
tensor is the same as that for the stiff fluid. The unit velocity vector, $u^\mu= (\dot{\phi}(t))^{-1}$   
In a comoving coordinate system, we can choose $u^\mu=(1,0,0,0)$, which for the massless 
scalar field would be equivalent to choosing $\dot{\phi}(t)=1$. If $\dot{\phi}(t)$
does not diverge, such a choice is always possible.

\section{Einstein equations, regularity and boundary conditions}

The dynamic evolution of the initial data, as specified on a 
spacelike surface of constant time is determined by 
the Einstein equations. For the metric (\ref{metric}), 
using the definitions 
\be
G(t,r)=e^{-2\psi}(R^{\prime})^{2}, H(t,r)=e^{-2\nu} (\dot{R})^{2}, 
\ee
and
\begin{equation}
F=R(1-G+H),
\label{eq:ein4}
\end{equation}
the independent Einstein equations for the 
massless scalar field (in the units $8\pi G=c=1$) are then 
given by,
\begin{equation}
F^{\prime}= \frac{1}{2}e^{-2\nu}\dot{\ph}^2R^{2}R^{\prime}\;, 
\label{eq:ein1}
\end{equation}
\be
\dot{F}=-\frac{1}{2}e^{-2\nu}\dot{\ph}^2R^{2}\dot{R} \;,
\n{eq:ein1a}
\ee
\begin{equation}
\partial_t\left(R^2e^{\psi-\nu}\dot{\ph}\right)=0\;,
\label{eq:ein2}
\end{equation}
\begin{equation}
R'\dot{G}-2\dot{R}\nu'G=0\; .
\label{eq:ein3}
\end{equation}
Here $(')$ denotes the partial derivative with respect to the 
coordinate $r$ and $(\dot{})$ with respect to $t$. 
The function $F=F(t,r)$ has an interpretation of the mass 
function for the collapsing cloud, and it gives the total mass 
in a shell of comoving radius $r$ on any spacelike slice 
$t=const$. The energy conditions imply $F\ge0$.

The function $R(t,r)$ is the area radius of a shell labeled 
$r$ at an epoch $t$.
For the sake of definiteness let us consider the situation 
of a collapsing cloud, and we have $\dot R<0$ as we are considering
the collapsing branch of the solutions. If $\dot R$ changes
sign then that corresponds to a bounce or dispersal of the field
during evolution.
We use the scaling freedom for the radial coordinate
$r$ to write $R=r$ at the initial epoch $t=t_i$, and in order to
distinguish the regular center of the cloud at $r=0$ from the genuine 
spacetime singularity at the termination of
collapse, where the area radius $R=0$ in both cases, 
we introduce a function $v(t,r)$ as defined by
\begin{equation}
v(t,r)\equiv R/r.
\label{eq:R}
\end{equation}
We then have,
\begin{equation}
R(t,r)=rv(t,r),\; v(t_i,r)=1,\; v(t_s(r),r)=0
\end{equation}
with $\dot{v}<0$.
The time $t=t_s(r)$ here corresponds to the shell-focusing
singularity at $R=0$, where the matter shell labeled a comoving radius
of constant $r$ collapses to a vanishing physical radius $R$ on  
reaching the genuine spacetime singularity.

We note that \e \eq{eq:ein2} is the Klein-Gordon 
equation for the scalar field, which is a part here of the Einstein 
equations via the Bianchi identities. We can integrate this equation 
to get
\be
R^2e^{\psi-\nu}\dot{\ph}=r^2\f,
\n{kg}
\ee
where $\f$ is an arbitrary function of integration.
We can now eliminate the function $\dot{\phi}(t)$ from equations 
\eq{eq:ein1} and \eq{eq:ein1a} to get
\be
\frac{F'}{R'}=-\frac{\dot{F}}{\dot{R}}=\frac{1}{2}
\frac{r^4f^2(r)G}{R^2R'^2}\;.
\n{ein6}
\ee
We now have four Einstein equations, namely \eq{eq:ein4}, \eq{eq:ein3} 
and \eq{ein6}, and four unknown functions of two variables, 
$\psi,\nu,R$ and $F$.
Solution of these equations, subject to the initial data and energy 
conditions, would determine the time evolution of the system.

Our purpose now is to construct the classes of solutions to the 
Einstein field equations, which give the dynamical  
scalar field collapse evolutions, given the initial data at 
an initial time $t=t_i$. We define the suitably differentiable functions 
$\M$ and $A(r,v)$ as below
\begin{equation}
\M\equiv \frac{F(t,r)}{r^3}\;,
\label{eq:mass}
\end{equation}
\begin{equation}
A(r,v)_{,v}\equiv \frac{\nu'}{R'}\;.
\label{eq:A}
\end{equation}
We note that Einstein equations \eq{eq:ein1} and \eq{eq:ein1a}, 
imply that $F$ must behave necessarily as $r^3$ closer to the 
regular center $r=0$ of the cloud, in order to preserve the regularity
of initial data and to preserve the finiteness of matter density
at all regular epochs of evolution. Hence, as $\M$ is a general, at least 
$C^2$ function, the \e(\ref{eq:mass}) is not really any ansatz 
or a special choice, but a fully generic class of mass profiles 
for the collapsing cloud, consistent with and as 
allowed by the regularity conditions.

Also, in order to be specific, we construct only
classes of collapse evolutions which admit no shell-crossing 
singularities in the spacetime where $R'=0$. We therefore consider 
only the genuine singularity at $R=0$ where the physical
radii of the collapsing shells vanish, and not the cases when 
nearby shells of matter may cross, giving 
rise to a density singularity which need not be gravitationally strong.
Therefore the function 
$A(r,v)$ is well-defined for all non-singular epochs.
Now using \e (\ref{eq:A}) in \e (\ref{eq:ein3}) we get,
as a class of solutions of Einstein's equations
\begin{equation}
G(r,v)=b(r) e^{2rA(r,v)}\;.
\label{eq:G}
\end{equation}
Here $b(r)$ is another arbitrary function of the shell radius $r$. 
The regularity condition on the velocity function $\dot{v}$ at the
center of the cloud implies that the form of $b(r)$ has to be,
\begin{equation}
b(r)=1+r^2b_0(r).
\label{eq:veldist}
\end{equation}
A comparison with the Lemaitre-Tolman-Bondi dust collapse 
models 
\cite{JD}
implies that we can interpret $b_0(r)$ as the energy distribution 
function for the collapsing shells.

We emphasize that the functions
$\M$ and $A(r,v)$ are not independent here in the case of scalar field
collapse. Because, using 
\eq{eq:G} in the second part of \eq{ein6} gives the required 
relation between them as
\be
2rA(r,v)=\ln\left[\frac{-2\M_{,v}v^2(v+rv')^2}{q(r)}\right]\;.
\n{am}
\ee
where $q(r)=f^2(r)[1+r^2b_0(r)]$. Now to determine the function $\M$, 
we use \eq{eq:mass} in the first part of \eq{ein6} 
to get the required first order equation
\begin{equation}
3\M + r\M_{,r} + Q(r,v)\M_{,v}=0 \;. 
\label{eq:eos}  
\end{equation}
where $Q(r,v)=(2rv'+v)$.
The above equation has a general solution of the form
\begin{equation}
{\cal F}(X,Y)=0,
\label{eq:solution}
\end{equation}  
where $X(r,v,\M)$ and $Y(r,v,\M)$ are the solutions
of the system of equations,
\begin{equation}
\frac{d\M}{3\M}=\frac{dr}{r}= \frac{dv}{Q} \;.
\label{eq:auxilliary}
\end{equation}   
Amongst all the classes of solutions $\M$ of above,
those are to be considered which obey the energy conditions
and required regularity conditions for the collapse.

To specify the boundary conditions for the cloud, 
solving \eq{eq:eos} at $r=0$, we get
\be
\lim_{r\rightarrow 0}\M=\frac{m_0}{v^3}\;,
\n{bc2}
\ee
where $m_0$ is a constant and from \eq{am} we see that the regularity 
at the center of the cloud requires $6m_0=q(0)$. 
Along the singularity curve $v=0$ the density diverges.
To solve for the function $v(t,r)$, we use the equation 
of motion \eq{eq:ein4}, and defining a function $h(r,v)$ as
\begin{equation}
h(r,v)=[e^{2rA(r,v)}-1]/r^2, 
\label{eq:h}
\end{equation}
we get,
\be
\sqrt{v}\dot{v}=-\G(r,v)
\n{vdot}
\ee
where,
\be
\G(r,v)=e^{\nu(r,v)}\sqrt{vb_0e^{2rA}+vh(r,v)+\M}.
\n{vdota}
\ee
The negative sign in the right hand side of \e \eq{vdot} 
corresponds to a collapse scenario where we have $\dot{R}<0$. 
Also in the $(r,v)$ plane, the function $\nu(r,v)$ is related to the 
function $A(r,v)$ by the relation
\be
\nu(r,v)_{,r}+\nu(r,v)_{,v}v'=A(r,v)_{,v}(v+rv')
\n{nurv}
\ee
Given the functions $A(r,v)_{,v}$ and $v'$ in terms of $r$ and $v$, 
this is again a quasi-linear first order partial 
differential equation like \eq{eq:eos}, with a similar general solution.

Now we have derived the relations between all the functions in the 
$(r,v)$ plane. The solutions to this system describes the evolving
scalar fields which we discuss, and then we consider the nature 
of initial data required to completely specify the collapsing model.


To see the existence of solutions to the complete system 
evolving from an initial spacelike slice $t=t_i$, consider 
the partial derivatives of the function $v(t,r)$.  
We get $v'$ from \eq{eq:eos}
\be
v'= - \frac{3\M + r\M_{,r} + v\M_{,v}}{2r\M_{,v}}\;.
\n{vdash1}
\ee
and from \eq{vdot} we get $\dot{v}$.
\be
\dot{v}=-\frac{\G(r,v)}{\sqrt{v}}\;.
\n{vdot1}
\ee
These give expressions of $\dot{v}$ and $v'$ in terms of $v,r$, 
$\M$ and its derivatives. As we have already seen 
from \eq{am}, we can write the function $A(r,v)$ and hence 
$\nu(r,v)$ in terms of $\M$ and it's derivatives, $r,v$ and $v'$. 
Thus, to get a solution of $v(t,r)$, which would in turn provide the 
complete dynamical solution to the collapsing system, we need the 
Pfaffian differential equation, $ v'dr+\dot{v}dt-dv=0$, to be integrable.
The integrability condition is given by,
\be
\dot{v}v'_{,v}=\dot{v}_{,r}+v'\dot{v}_{,v}
\n{intcond}
\ee
This integrability condition gives the required second order 
equation for the mass function $\M$. 
Any solution of the above equation 
would by default solve the quasilinear equation \eq{eq:eos}, via \eq{vdash1}.
Also, this would then uniquely solve for functions $A(r,v)$ and hence 
$\nu(r,v)$ via \eq{am}. 
The solution set of this integrability condition is non-empty as 
$\M=m_0/v^3$ solves the whole system with $v=v(t)$, $\nu=\nu(t)$ 
and $A=A(r)$ to give a {\it FRW} interior.

For each solution $M$ of 
\eqref{intcond}, we would have a particular Pfaffian equation, which would be 
integrable. We know that given one 
integrating factor for a Pfaffian differential equation, we can find 
infinity of them. So for any such $M$, if we have one such 
integrating factor of the corresponding Pfaffian equation, then there would be 
an infinite number of such integrating factors. In general, they would correspond 
to an infinite number of solutions of that Pfaffian equation. 
Among the solutions that exist, we choose 
only those which obey the required boundary conditions.

We now consider the independent initial profiles required at the 
epoch $t=t_i$ or $v=1$ to evolve the system via Einstein
equations. From \eq{eq:ein1} we see that providing the 
function ${\cal M}(r,1)$, which is the same as $F(t_i,r)/r^3$, 
would determine the function $\nu(t_i,r)$ (upto a multiplicative 
constant $\dot{\ph(t_i)}$). This would then determine the function 
$A(r,v)_{,v}|_{v=1}$ using \eq{eq:A} or \eq{nurv}. 
Also from \eq{eq:eos}, we see 
that ${\cal M}(r,1)$ specifies $\M_{,v}|_{v=1}$, which in turn 
determines $A(r,1)$ using \eq{am} and $h(r,1)$. 
Hence we see that specifying the functions ${\cal M}(r,1)$ (interpreted 
as the initial density profile) and $q(r)=f^2(r)[1+r^2b_0(r)]$ (which 
specifies the initial velocity profile) completely determine 
the evolution of the system.

Assured of the existence of solutions, integrating \e \eq{vdot} 
with respect to $v$, we can write the solution of $v(t,r)$ 
in an integral form,
\begin{equation}
t(v,r)=\int_v^1\frac{\sqrt{v}dv}
{\G(r,v)}\;.
\n{tvr1}
\end{equation}
Note that the variable $r$ is treated as a constant in the above 
equation. The above gives the time taken for a shell labeled 
$r$ to reach a particular epoch $v$ from the initial epoch $v=1$.

One could sketch the iterative process by which the 
initial Cauchy data evolves. As we noted, given the 
density and energy profile we can find all the other functions at 
the initial epoch $v=1$. Then we use \eq{tvr1} to find the 
functional form of $v(t,r)$ for $v=1-\epsilon$ where $\epsilon$ is 
an infinitesimally small positive number. As we easily see, this 
functional form would be
\be
v(\Delta t, r)= 1-j(r)\Delta t,
\ee
where 
\be
j(r)=e^{\nu_0(r)}\sqrt{b_0e^{2rA_0(r)}+h_0(r)+{\cal M}_0(r)}.
\ee
Here the subscript $0$ denotes the initial data and $\Delta t$ 
denotes the infinitesimal evolution of the spacelike initial slice 
in $(t,r)$ plane.
We now calculate the function $v'$ as
\be
v'=j'(r)\Delta t= j'(r)(1-v)/j(r)
\ee
Using this in \eq{eq:eos} gives 
$\M$ at $v=1-\epsilon$ with respect to the given initial data 
${\cal M}_0(r)$ and boundary conditions. Using the form of 
$\M$ and $v'$ we then calculate the function $A(r,1-\epsilon)$ 
using \eq{am}, $\nu(r,1-\epsilon)$ using \eq{nurv}, and finally 
$h(r,1-\epsilon)$. We then plug all these functions 
in \eq{tvr1} again for the next iteration, until we reach $v=0$.

Now we see that the time taken for a shell labeled 
$r$ to reach the spacetime singularity at $R=0$ (which is the 
{\it singularity curve}), is given as
\begin{equation}
t_s(r)=\int_0^1\frac{\sqrt{v}dv}
{\G(r,v)}\;.
\n{tsr1}
\end{equation}
In a physically realistic gravitational collapse situation
such as collapse of a massive matter cloud which continually collapses, one
would focus only on those classes of solutions where $t_s(r)$ is finite and 
sufficiently regular. This means that the cloud collapses in a finite 
amount of time.

\section{Collapse of the scalar field to a simultaneous singularity}

We now consider the case when the singularity occurring in the 
spacetime as a result of collapse is simultaneous. This would be the 
case, for example, when the density is same at every point in space 
at any given time, i.e. the collapse is homogeneous. In this case, 
$\rho=\rho(t)$. Since $\rho=e^{-2\nu}\dot{\phi}(t)^2$, therefore
we have $\nu=\nu(t)$ only. We rescale $t$ to make $e^{2\nu}=1$. 
The spacetime singularity 
occurs when the physical radius goes to zero, i.e. $v=0$. In the 
case of the density being homogeneous, this implies that 
the singularity curve $t_s(r)$ is independent of $r$.

The expression for $t_s(r)$ in general is, 
\begin{equation}
 t_s(r)=\int_0^1 \frac{\sqrt{v}}{[\frac{v}{r^2}(G-1)+M]^{\frac{1}{2}}} 
dv.\label{singcurv}
\end{equation}
Here the $r$-independence of $t_s(r)$ then implies that 
the integrand on the right hand side is a function of $v$ only.
The time coordinate can be expressed in general as 
\begin{equation}
t=\int \frac{\sqrt{v}}{[\frac{v}{r^2}(G-1)+M]^{\frac{1}{2}}} dv 
+h_1(r)\label{tcoord}
\end{equation}
where $h_1(r)$ is an arbitrary function of $r$. As the integral is  
a function of $v$ only, so the initial condition which is $v=1$ at 
the time $t=t_i$, implies that $h_1$ is a constant. 
Therefore from \eqref{tcoord}, it is seen that 
\begin{equation}
 v'=0 .
\end{equation}

>From the Einstein equations, we get the relation
\begin{equation}
 e^{-2\nu}\dot{\phi(t)}^2=-\frac{2M,_v}{v^2}\label{phi}
\end{equation}
Since the left hand side of the equation is a function of 
$t$ only, the right side must be a function of $v(t)$ only. This 
implies $M=M(v)$ only. So \eqref{eq:eos} gives 
\begin{equation}
 M=\frac{m_0}{v^3}
\end{equation}
>From \eqref{eq:A} we get, 
\begin{equation}
 A,_v=0
\end{equation}
This implies $A=A(r)$. This is consistent with the form of $G$ 
we get from the other equation,
\begin{equation}
G=-\frac{2v^2M,_v(v+rv')^2}{f^2(r)}\label{Gequn}
\end{equation}
Putting $M$ in the last equation, we get $G=\frac{6m_0}{f^2(r)}$. 
Since $t_s(r)\neq 0$, the integrand in its expression must be finite 
at $r=0$. This implies,
\begin{equation}
\frac{1}{r^2}(\frac{6m_0}{f^2(r)}-1)=f_1(r)
\end{equation}
where $f_1(0)$ is finite. In this case, since $t_s(r)$ is a constant, 
$f_1(r)$ is a constant also. So we can write 
\begin{equation}
f^2(r)=\frac{6m_0}{1+cr^2}
\end{equation}
where $c$ is a constant.
So in this case we have $e^{2\psi}=\frac{v^2}{1+cr^2}$. 
>From \eqref{tcoord} we get,
\begin{equation}
 t=-\int \frac{\sqrt{v}}{(cv+\frac{m_0}{v^3})^{\frac{1}{2}}} dv
\end{equation}
The metric then is given as, 
\begin{equation}
 ds^2= dt^2- v(t)^2(\frac{dr^2}{1+cr^2}+r^2d\Omega^2)
\end{equation}

In this case, $t_s(r)$ is finite and $\dot{\phi(t)}$ blows up 
at the singularity. This can be seen from \eqref{phi} as it reduces 
to $\dot{\phi(t)}^2= \frac{6m_0}{v^6}$. We also note that $c=0,\pm1$, 
gives all the possible solutions in this case, which correspond to 
the Friedmann-Robertson-Walker models. 
When $c=0$, we have $v(t)= (K_1-3 \sqrt{m_0}t)^{\frac{1}{3}}$, 
where $K_1$ is some positive constant.
 
We note here that we considered above the case when the 
density $\rho= \rho (t)$, which leads to a simultaneous singularity,
and the FRW class of models. However, even when the density 
is not homogeneous, a simultaneous singularity can result 
as we shall discuss below.

\section{Collapse of a scalar field of inhomogeneous density}

In this case, the singularity that results as the collapse 
endstate need not be simultaneous in general, and in general we have 
$\rho=\rho(r,t)$. Before proceeding further, we discuss the regularity  
conditions that are required for a physically reasonable collapse 
of the scalar field.

\subsection{Regularity conditions}

First, we note that at the center when $r$ goes to zero, we 
must have
\begin{equation}
 \lim_{r \to 0}(rv')=0 \label{rvprime}
\end{equation}
Because, when this condition is violated, $v$ becomes divergent as $r$ 
goes to zero at the center. For example, let $\lim_{r \to 0}(rv')=c_2$, 
where $c_2$ 
has some non-zero value. Then $\lim_{r \to 0}v$ goes 
as $\mod{lnr}$. If $\lim_{r \to 0}(rv')$ is divergent in $r$, then the
divergence of $v$ will only become more severe. But divergence of $v$
means that $R'$ becomes arbitrarily large for the
comoving shells near the center. This is clearly unphysical and 
is to be ruled out.

The second condition comes from one of the Einstein equations. 
In the limit of going to the center $\lim{r \to 0}$, using \eqref{rvprime} 
in \eqref{eq:eos}, we get
\begin{equation}
 3M + rM,_r + vM,_v =0 \label{eqeos1}
\end{equation}
Here $\lim_{r \to 0}(rM,_r)$ can behave in three possible ways, 
which are, i) $\lim_{r \to 0}(rM,_r)=0$, (ii) $\lim_{r \to 0}(rM,_r)$ has 
some finite non-zero value, and (iii) $\lim_{r \to 0}(rM,_r)$ can be 
divergent.

Now we point out that only the first option is possible for a 
physically reasonable collapse model. In the case of (ii) occurring, we have 
$\lim_{r \to 0}M= c_0 (lnr)+h_2(v)$. Now we can argue that $h_2(v)$ is
not divergent. Otherwise, $M,_v$ will also contain a logarithmic
divergence. But for $v\neq 0$, this implies a divergence of density,
which is not physical. So we cannot consider that case. For the other
case, one can see from \eqref{singcurv}, that the integrand becomes
divergent when $r$ goes to zero. Therefore $t_s(0)=0$, which means
that the singularity is present even at the initial epoch. So (ii) is 
ruled out. Similarly, one can also rule out the possibility (iii), 
proceeding in the same way as for the case (ii). In this case also 
$M$ has $r$-dependent divergence, which makes $t_s(0)=0$. Therefore,  
(i) is the only possibility that is allowed.\\

In such a case then, \eqref{eqeos1} reduces to $3M+vM,_v=0$ in 
the $\lim{r \to 0}$. This implies $\lim_{r \to 0}M(r,v)= m_0/v^3$, 
for $v\neq0$.

\subsection{Black hole formation in scalar field collapse}

In the following, we consider and characterize a wide 
class of black hole models that arise in the gravitational collapse 
of a massless scalar field, and which satisfy the regularity 
conditions above. It is shown that for a large class of collapse
scenarios the field collapses to a simultaneous singularity.

We note that the second condition that we adopted above can 
be stated as
\begin{equation}
 M(0,v)=\frac{m_0}{v^3} \label{Mcentre}
\end{equation}
for all $v\geq0$, i.e. $1\ge v \ge 0$.
At this point, it is useful to note that $\phi(t)$ is a free 
function and therefore can be chosen to be any function of $t$ subject 
to its being regular. However, once the regularity conditions have been
imposed, then two solutions (obeying the same regularity conditions),
with two different functional forms of $\dot{\phi}(t)$ need not be
diffeomorphic to each other in general.

We note that any $r=constant$ curve is timelike, and the tangent 
vector of this curve is, $\tau^\mu= dx^\mu/ds$ with components,
\begin{equation}
\tau^\mu=(\frac{dx^0}{ds},0,0,0) 
\end{equation}
The proper time along this curve is then given by
$\tau=\int \tau^\mu\tau_\mu ds=\int ds$. Since $ds^2=e^{2\nu}dt^2$, 
in this case we get $\tau=\int e^\nu dt$, and we have,
\begin{equation}
 \tau[v(t_f),r]= \int_{v(t_i)}^{v(t_f)} e^\nu dt
\end{equation}
So $\tau(v,r)$ is the proper time along any particular shell 
of comoving radius $r$ to reach the value $v$, starting from the initial 
epoch $v=1$.

Subject to the above mentioned conditions, some general results can 
now be proved about the nature of singularities and black hole formation 
in gravitational collapse of scalar fields. 
We thus consider the classes of
models where $\dot{v}\leq0$ through out the evolution. 
Further, we take that $v'\geq 0$ at $r=0$, and at all other values 
of $r$. As we show here, this last condition would imply
that this class of solutions would admit no non-simultaneous 
singularities as collapse endstate. 
Such a condition corresponds to the situation when the central 
shell at $r=0$ arrives at the spacetime singularity earlier in time 
as compared to other shells with greater values of $r$. This is 
related to avoiding the shell-crossing singularities $R'=0$ within the 
cloud which are not generally considered to be physically genuine. 
We shall show that, in such a case, the matter field then collapses to 
either a simultaneous singularity in a finite coordinate time, or 
that both $t_s(r)$ and the proper time $\tau(v,r)$ must diverge along 
any timelike curve $r= constant$ as the system evolves
under the conditions we shall state below.

{\bf Proposition 1:} If $\dot{\phi}(t)$ is divergent at some instant 
$t_1$, then there is a simultaneous singularity at the time $t=t_1$.\\

{\bf Proof:} The density $\rho= \frac{1}{2}e^{-2\nu}\dot{\phi}(t)^2$. If 
$\dot{\phi}(t)$ is divergent, at $t=t_1$, then there are two possibilities.
The first possibility is, the density remains finite and there is no 
singularity at $t=t_1$ in the case if we have
$e^{-2\nu(r,t_1)}=0$. However, this is not allowed because it means   
$e^{2\nu(r,t_1)}=\infty$, i.e. the
metric component is blowing up at regular spacetime points. As this 
is not allowed by regularity conditions, the only other possibility 
that remains is the density must diverge at $t=t_1$, so the singularity 
is simultaneous at the epoch $t=t_1$.\\

%

Therefore, it follows that the existence of a non-simultaneous 
singularity curve implies that the function $\dot{\phi(t)}$ must remain 
finite at all regular epochs in the spacetime.
For a stiff fluid collapse, this means that if $\lim{u^0 \to 0}$ 
as $\lim{t \to t_s}$, there must be a simultaneous singularity 
in the limit of $t$ going to $t_s$.

%

{\bf Proposition 2:} If $t_s(r)$ is not constant and if $v'\ge0$
everywhere in the spacetime, then $t_s(r)$ must be divergent.\\

{\bf Proof:} First we note that from \eqref{Gequn} we get, 
\be
G(r,v)=-\frac{v^2(vM,_v-3M-rM,_r)^2}{2f^2(r)M,_v}
\ee
Therefore,
\be
t_s(r)=\int_0^1 \frac{\sqrt{v}}{e^\nu
[\frac{v}{r^2}(-\frac{v^2(vM,_v-3M-rM,_r)^2}{2f^2(r)M,_v}-1)+M]^{\frac{1}{2}}}dv
\ee 
We note that $e^{\nu(0,v)}\neq0$ when $v\neq0$ by the 
regularity conditions. This implies that the other term in the denominator 
of the integrand must be finite at $r=0$, otherwise $t_s(0)=0$, or 
the singularity will be present at the initial epoch itself, which 
violates the regularity of collapse from non-singular initial data. 
We note that $M(0,v)=m_0/v^3$ for all values of $v$ such that 
$1\ge v \ge0$ as stated above, which is finite whenever 
$v\neq0$. Therefore the quantity,
\be
\lim_{r \to0}
\frac{1}{r^2}[-\frac{v^2(vM,_v-3M-rM,_r)^2}{2f^2(r)M,_v}-1]
=\lim_{r \to0}X 
\ee
must be finite. This finiteness condition and \eqref{Mcentre} 
together imply that $M(r,v)$ must be of the form,
$M(r,v) = \frac{m_0}{v^3}+ r^n g(r,v)$, where $n\geq2$. 
This can be clearly seen by the direct substitution of this form for
$M$ in $t_s(r)$ (see also Appendix A). Now we can write,
\be
\lim_{r \to0}X= \lim_{r \to0}
[\frac{1}{r^2}(\frac{6m_0}{f^2(r)}-1)+2r^{n-2}\frac{v^3}{f^2(r)}((n+3)g+g,_r))
+\frac{r^{2n-2}((n+3)g+g,_r+vg,_v)^2v^6}{2f^2(r)(3m_0-v^4r^ng,_v)}]
\ee
Now $\frac{1}{r^2}(\frac{6m_0}{f^2(r)}-1)\equiv f_1(r)$, 
where $f_1(0)$ is a finite quantity in the limit of $r\to0$
because $X$ cannot diverge. So we get,
\begin{equation}
f^2(r)=\frac{6m_0}{1+r^2f_1(r)}
\end{equation}
When $n>2$, we can see from above that $\lim_{r \to 0}X=f_1(0)$, 
and when $n=2$, $\lim_{r \to 0}X=f_1(0)+\frac{v^3}{f^2(0)}g_0(v)
=f_1(0)+\frac{v^3}{3m_0}g_0(v)$, where $g_0(v)= (n+3)g(0,v)+g,_r(0,v)$.\\

>From this, it follows therefore that in the case $n>2$, 
\begin{equation}
 \lim_{r \to 0}\lim_{v \to 0} vX(r,v)+M(r,v)= \frac{m_0}{v^3}
\end{equation}
When $n=2$, $O[g_0(v)]>O(\frac{1}{v^7})$ which violates
the continuity of $\tau_s(r)$ and hence is ruled out (see Appendix B).
Therefore we have,
\begin{equation}
 \lim_{r \to 0}\lim_{v \to 0} vX(r,v)+M(r,v)= \frac{c_1}{v^3}
\end{equation}
where $c_1$ is some constant, which is the result in general for
$n\ge2$.

We now note that from \eqref{phi}, we have
$e^{\nu(0,v)}=\frac{v^3\dot{\phi}(t)}{\sqrt{6m_0}}$.  
This can always be written as
\begin{equation}
 \lim_{v \to 0}e^{\nu(0,v)}= v^3 f_3(v) \label{nu0v}
\end{equation}
The divergence of $t_s(0)$ if it is there can come from the 
range where $v$ is close to zero. We denote that part of the
integral by $t_{sd}(0)$. So we can write, 
\be
t_{sd}(0)=\int_0^\epsilon
\frac{\sqrt{v}}{v^3f_3(v)(\frac{c_2}{v^3})^{\frac{1}{2}}} dv
\ee
Here $c_2=m_0$ or $c_2=c_1$, and $\epsilon$ is a small 
quantity. This can be written as,
\begin{equation}
t_{sd}(0)= \frac{1}{\sqrt{c_2}} \int_0^\epsilon \frac{1}{vf_3(v)} dv
\end{equation}
Now, if the singularity is non-simultaneous then 
$\dot{\phi}(t)$ remains finite always. Therefore, $f_3(v)$ is also 
finite. So we get the result that $t_s(0)$ is divergent. 
Since $v'\ge0$, it follows that $t_s(r)$ is also divergent.
This proves the required result.

We note that if the singularity is simultaneous, then 
$\dot{\phi}(t)$ blows up when $v(0,t)=0$ by Prop. 1. Then $f_3(v)$
must be divergent and then $t_s(0)$ has to be finite.
For a collapsing massless scalar field, if the mass function $\M$ 
and all the 
metric functions are at least $C^2$ near the central 
shell as they should be, and the singularity curve is non-simultaneous 
and an increasing function of time near the center, then the time 
taken for the central shell to reach the singularity diverges 
logarithmically, as we have shown above.

As an illustrative example, we can consider the class of 
solutions within 
an $\epsilon$-ball around the central shell, where we can
ignore the contributions of terms higher than $r^2$ even very close to
$v\to0$. We note that this may not be possible for all classes of 
solutions, as in the Einstein equations there are terms of the 
form $r^n/v^m$ ($m,n>0$) and these may have non-zero limit in 
the vicinity of the singularity, that is $v=0$.

Since the mass function is $C^2$ near the center, without any loss of
generality we can Taylor expand the function around $r=0$ as,
\be
\M=\frac{\mo}{v^3}+\mt r^2+\cdots
\n{mform}
\ee
We note that we get the zeroth order term by solving \eq{eq:eos} at 
$r=0$. Also that the first order term should vanish follows directly from 
the {\it no force condition} at the center of the cloud. The function 
$\mt\;$ is an unknown function which solves \eq{eq:eos} near the center.
Using \eq{mform} we can write \eq{vdash1} as
\be
v'=\frac{r}{6\mo}[v^5\mt]_{,v}\left[1+\frac{\mtv v^4}{3\mo}r^2+
\cdots \right].
\n{vdash4}
\ee 
Let us consider $b_0(r)=0$ and $\f=f(0)=6\mo$. Then putting the
Taylor expanded forms of different functions in the \eq{am} we get
\be
e^{2rA}=1+ \left(\frac{5v^3\mt}{3\mo}\right)r^2+{\cal O}(r^6)
\n{A}
\ee  
Also using \eq{h} we have
\be
h(r,v)=\left(\frac{5v^3\mt}{3\mo}\right)+{\cal O}(r^4).
\n{h}
\ee

>From the above equations we get the behaviour of the function 
$\nu$ near the center (since $\nu'=A_{,v}R'$) as
\be
\nu\approx \frac{5v}{12\mo}[v^3\mt]_{,v}r^2+{\cal O}(r^4)+ C(t), 
\n{nu}
\ee
where $C(t)$ is the constant of integration. Comparing with the densilty of 
the central shell we get $C(t)=ln(v^3/m_0)$.
Therefore we can write,
\be
e^\nu\approx \frac{v^3}{m_0}\left[ 1+\frac{5v}{12\mo}[v^3\mt]_{,v}r^2
+{\cal O}(r^4)\right]\n{nu1}
\ee
Finally using all the above equations and neglecting terms higher 
than the order $r$ (since we are only concerned with an $\epsilon$-ball 
around the center), we get,
\be 
\dot{v}=-\frac{v^3}{m_0}\sqrt{\frac{5v^3\mt}{3m_0}+\frac{m_0}{v^4}}\;;
\;v'=\frac{r}{6\mo}[v^5\mt]_{,v}
\ee
>From \eq{intcond} we see that the integrability condition is 
trivially satisfied for $r=0$. However, upto the first order correction 
we get the following condition,
\be
\left[(v^5\mt)_{,v}\right]^2=C\frac{v^6}{m_0^2}
\left[\frac{5v^3\mt}{3m_0}+\frac{m_0}{v^4}\right]
\ee
The above is a complicated non-linear equation for the function 
$\mt$. Let us consider that near the central singularity $v(0,t)=0$, the 
function has a leading order power law behaviour $\mt\approx v^\alpha$. 
Then using the above equation and comparing powers on both sides, 
we see that only consistant solution can be obtained with 
$\alpha=-3$. In other words the behaviour of the function \mt near the 
central singularity is of the order of $1/v^3$. Using this in 
the singularity curve expression \eq{tsr1} we see that the time 
taken for the central shell to become singular diverges logarithmically.

Returning to the class of collapse models under consideration, 
we shall now show that for this class non-simultaneous singularity cannot 
occur. First we define,
\begin{equation}
\tau_0(t)=\int_{t_i}^{t} e^{\nu(0,t)}dt \label{defn1}
\end{equation}
This is the proper time along the $r=0$ shell. Now let us assume that 
$\tau_0(t_{s0})=\tau_{0s}$ is finite.
>From \eqref{defn1} we get,
\begin{equation}
\frac{d\tau_0}{dt}=e^{\nu(0,t)}
\end{equation}
We can then show the following result.

{\bf Proposition 3} If $v'>0$ at $\tau_0=\tau_{0s}$, then for 
any $r_2>0$, $\tau_{r_2}(\tau_{0s})$ is divergent.\\

{\bf Proof:} One can make the following transformation 
which is allowed by the comoving coordinates, given by, 
$x^\mu: (t,r,\theta,\phi) \to x'^\mu=(\tau_0,r,\theta,\phi)$.
In the new coordinate system, the line element is 
then written as,
\begin{equation}
ds^2=e^{2[\nu(r,\tau_0)-\nu(0,\tau_0)]}d\tau_0^2-e^{2\psi}dr^2-R^2d\Omega^2
\end{equation}
The new coordinates cover the manifold upto $\tau_0=\tau_{0s}$ only.
Since, $v'>0$ at $\tau_0=\tau_{0s}$, $(r_2,\tau_{0s})$ is a 
regular spacetime point. The proper time along the $r_2=const.$ 
curve from $\tau_0=\tau_{0i}$ to $\tau_0=\tau_{0s}$ is given by, 
$\tau_{r_2}(\tau_{0s})=\int_{\tau_{0i}}^{\tau_{0s}} \sqrt{g_{00}'
(r_2,\tau_0)} d\tau_0= \int_{\tau_{0i}}^{\tau_{0s}}e^{\nu(r_2,\tau_0)
-\nu(0,\tau_0)} d\tau_0$.\\

Now $v(0,t)=v_0(t)$ is a function of $r$ only. Therefore 
$dt=\frac{1}{\dot{v}(0,t)}dv_0=\frac{1}{\dot{v}(0,v_0)}dv_0$.
This implies, 
\begin{equation}
d\tau_0=\frac{e^{\nu(0,v_0)}}{\dot{v}(0,v_0)}dv_0
\end{equation}
Using this, we get 
\begin{equation}
\tau_{r_2}(\tau_{0s})= \int_{v_0=1}^{v_0=0}\frac{e^{\nu[r_2,v_0]}}
{\dot{v}(0,v_0)}dv_0 
\end{equation}
Here it is important to note that $e^{\nu[r_2,v_0]}
=e^{\nu[r_2,v(r_2,v_0)]}$ is finite and non-zero at $v_0=0$, because
$v(r_2,0)>0$.\\

>From the proof of Prop. 2, we have $\lim_{v_0 \to 0} \dot{v}(0,v_0)
=-c_1v_0$, where $c_1>0$. Since $\dot{v}$ goes to zero as $v_0 \to 0$, a 
divergence may be present in $\tau_{r_2}$. So we consider that part 
of the integral where $v_0<<1$, which is given by,
$\tau_{r_2}(\tau_{0s})_d=\frac{1}{c_1}\int_0^\epsilon
\frac{e^{\nu[r_2,v_0]}}{v_0}dv_0$, where $\epsilon<<1$.
This gives $\tau_{r_2}(\tau_{0s})_d=\frac{e^{\nu[r_2,v_0]
\mid_{v_0=0}}}{c_1}\int_0^\epsilon \frac{1}{v_0}dv_0$, which is  
a divergent quantity. It follows that $\tau_{r_2}(\tau_{0s})$ 
is divergent.

The proposition above proves that any non-simultaneous 
singularity cannot develop for this class of collapse models. That 
implies for this class only a spacelike singularity can form
as collapse endstate. This characterizes a wide class of black hole
formation models from a massless scalar field collapse, or for a
stiff fluid collapse.

We now also indicate here the possibility of existence of a 
class of non-singular solutions in evolving scalar fields,
or stiff fluids scenario, by proving another proposition as below.

{\bf Proposition 4:} If $v'(r,t)\geq b$ where $b>0$, for $r_1\leq r\leq r_2$ 
for some $r_1, r_2>0$, and $t\in (t_i,\infty)$, then we must have 
$\tau(v_1,r)>k$ for all $k>0$ for all $r$, for some $v_1>0$.\\

{\bf Proof:} For $r\geq r_2$, we have
$\int_0^r v'(r,t)dr = \int_0^{r_1} v'dr + \int_{r_1}^{r_2} v'dr 
+ \int_{r_2}^{r} v'dr$.\\
This implies that, $\int_0^r v'(r,t)dr\geq b(r_2-r_1)$ for $r\geq r_2$.\\
$\int_0^r v'(r,t)dr= v(r,t)-v(0,t)$\\ 
Denoting $b(r_2-r_1)$ by $c$,\\
$v(r,t)\geq v(0,t)+c$; for $r\geq r_2$.\\
$\because  v(0,t)\geq 0$ for $t\in (t_i,\infty)$, therefore 
$v(0,t)\geq c$ for $t\in (t_i,\infty)$ for $r\geq r_2$.\\
For $r\geq r_2$, the proper time elapsed during the comoving time
interval $(t_i,\infty)$, $\tau(v_1,r)= \int_{t_i}^\infty e^\nu dt$, 
for some $v_1\geq c$.\\
By the regularity condition, $e^{\nu(r,v)}$ has a positive lower 
bound in the range $v\in (c,1)$.\\
This implies that, $\tau(v_1,r)>k \forall k>0$ for $r\geq r_2$ 
and for some $v_1\geq c$.\\
Since $\tau(v_1,r)$ is continuous in each $t= constant$ hypersurface,
and $r_2$ can be any non-zero number however small, there must exist
some $v>0$ such that  $\tau(v,r)>k \forall{k}>0$  for $r<r_2$. 
So $\tau(v,r)>k \forall {k}>0$ for $\forall r$ for some $v>0$.
This proves the result.

Thus all the classes of solutions that satisfy the conditions 
assumed as above would be singularity free. If there are no such solutions, 
then that would indicate 
the possibility of bouncing models within the framework above.
We note that since singularities would not form in this class of models, 
there would not be any formation of trapped surfaces also. 
We discuss this in some detail below.

For spherically symmetric spacetimes, the equation of the 
apparent horizon is given by
\begin{equation}
 g^{\mu\nu}R,_\mu R,_\nu=0
\end{equation}
In this case, this implies $G-H=0$. Using \eqref{eq:ein4}, one can 
rewrite the condition for formation of apparent horizon as $F=R$ or as
\begin{equation}
 M=\frac{v}{r^2} \label{apphn}
\end{equation}



{\bf Proposition 5:} For the non-singular class of models discussed 
in Proposition 3, the apparent horizon does not form in any finite coordinate 
time.\\

{\bf Proof:} To prove this, we first prove another result.\\

{\bf Lemma A:} We must have $\mod{\dot{v}}<\delta$, $\forall \delta>0$, 
in a total time interval, which is infinite.\\

{\bf Proof:} We know, $t_s(r)=\infty$, $\forall r$. Now we have,
\begin{equation}
 v(r,t_s(r))-v(r,t_i)=\int_{t_i}^{t_s(r)} \dot{v}dt.
\end{equation}
This can be written as $1=\int_{t_i}^\infty (-\dot{v})dt=
\int_{t_i}^{t_1} (-\dot{v})dt + \int_{t_1}^{t_2} (-\dot{v})dt+.....
=I_1+I_2+....$, where $(t_1-t_i)$, $(t_2-t_1)$, ....are all finite.\\
This implies that, given any $\delta_1>0$, there exist infinite number 
of integrals $I_n$s such that, $I_n<\delta_1$.\\
In each of these integrals, $\mod{\dot{v}}<\delta$, 
$\forall \delta>0$; i.e. mod(change of $v$)$<$ any arbitrarily small 
amount. Since there are infinite number of such integrals, 
$\mod{\dot{v}}<\delta$, $\forall \delta>0$, in a total time interval 
that is infinite.(Proved)

Since $\mod{\dot{v}}<\delta$, $\forall \delta>0$, for a 
total time interval which is infinite, there must exist a time 
instant $t_1$ at which this holds, such that, $t_1<t_{s0}$ and 
$t_1>l$ for any given $l>0$.\\
$\because t_s(r)\geq t_{s0}$, $\forall r$, so for some $v>0$, 
$v(r,t_1)>0$ for $r>0$. By the regularity condition, at $t=t_1$, 
$e^{\nu(r,t_1)}>0$. From Einstein equations we have $\sqrt{v}\dot{v}
=-e^\nu[\frac{v}{r^2}(G-1)+M]^{\frac{1}{2}}$.\\
Therefore at $t=t_1$,
\begin{equation}
[\frac{v}{r^2}(G-1)+M]<\delta, \forall \delta>0 . 
\end{equation}
By the energy condition, $\rho=-\frac{M,_v}{v^2}>0$. This 
implies $M,_v<0$. From \eqref{Gequn}, $G>0$.\\
Let, $v(t_1,r)=v_f(r)$. Then the inequality can be written as 
$M<\delta+(1-G)\frac{v}{r^2}$, $\forall \delta>0$. But we know, 
$(1-G)\frac{v_f}{r^2}<\frac{v_f}{r^2}$. This implies
\begin{equation}
 M(r,v_f)<\frac{v_f}{r^2}.
\end{equation}
Let $v_e(r)$ be the value of $v(r)$ at any time earlier than $t_1$.\\
$\frac{v_f(r)}{r^2}\leq \frac{v_e(r)}{r^2}$(Since $\dot{v}\leq 0$). 
Also $M(v_f,r)>M(v_e,r)$. These imply $M(v_e,r)<\frac{v_f(r)}{r^2}\leq 
\frac{v_e(r)}{r^2}$ or $M(v_e,r)<\frac{v_e(r)}{r^2}$.\\
Therefore the equation of apparent horizon \eqref{apphn} is not 
satisfied for $r\neq 0$ and when $t_i\leq t\leq t_1$. So there is no 
apparent horizon and consequently no trapped surface formation in 
that time range. Since there always exists some $t_1$ such that 
$t_1>l$ for any given $l$, for $r\neq0$, trapped surface does not 
form in any finite time.\\

For $r=0$, $M=\frac{m_0}{v^3}$. Putting this in \eqref{apphn}, 
$m_0r^2=v^4$. This is satisfied when $v=0$, i.e. at $t=t_{s0}$. Since  
$t_{s0}= \infty$, apparent horizon does not form at the $r=0$ shell 
at any finite time. So for such a solution of the Einstein equations, the 
apparent horizon or 
trapped surfaces do not form at any finite coordinate time.(Proved)\\

In Proposition 4, it has already been shown that an infinite 
coordinate time interval correspond to infinite proper time interval 
along any $r=constant$ world line. This implies that for this class 
of models, no apparent horizon or trapped surface would form at any 
finite proper time.

\subsection{Models collapsing to form non-simultaneous singularity}

Now we discuss briefly the other classes of models for which 
the singularity can be non-simultaneous. Toward such a purpose, we 
must clearly relax the condition $v'\geq 0$, thus allowing $v'$ 
to be either positive or negative. All the other regularity conditions 
still remain the same. 

In the earlier class of models, the divergence of $t_s(r)$ 
was responsible for the fact that there could only be simultaneous 
singularity. The divergence of $M$ as $v \to 0$ caused the divergence 
of $t_s(r)$. So one must look for such a form of $M$ which is 
different in this respect. One such generic form of $M$ is given by 
\be
M=\frac{m_0}{v^3}e^{-r^n g_1(r,v)}+r^2 g_2(r,v),
\ee
where $n\geq2$, $g_1(r,0)$ blows up and $g_2(r,0)$ is finite. 
In this case $M(r,0)$ would remain finite except at $r=0$.

Here $t_s(0)$ will be divergent as in the previous case, 
because $M(0,v)=\frac{m_0}{v^3}$. But in this case, $t_s(r)$ would be 
finite for any $r>0$. To show this we first show that there is no 
divergence in $t_s(r)$ when evaluating the integral near $v=0$. 
We know that $\lim_{v \to 0}M(r,v)=r^2g_2(r,v)$, where $g_2$ is
a regular function. Also, $\lim_{v \to 0}M(r,v),v= r^2g_2(r,v),v$ and 
\be
\lim_{v \to 0}v'= -\frac{5g_2(r,0)+rg_2(r,0),r}{2rg_2(r,v),v\mid_{v=0}}
\ee
>From \eqref{Gequn}, it is seen that $\lim_{v \to 0}G\sim v^2$ 
and from \eqref{phi}, $\lim_{v \to 0}e^{\nu(r,v)}\sim v$. Putting them 
all in \eqref{vdot}, we have $\lim_{v \to 0}\dot{v}\sim -\sqrt{v}$. 
This means that there would be no divergence in the $t_s(r)$ integral 
for any value $r>0$ coming from the part near $v=0$. So $t_s(r)$ 
is finite for $r>0$. We note that since $t_s(0)=\infty$ and $t_s(r)$ 
for $r>0$ is finite, $v'\geq0$ is not satisfied. The proper time for 
the $r=0$ shell to reach the singularity is, however, finite as we 
can show. The proper time, $\tau_{s0}=\int_{t_i}^{t_{s0}} e^\nu dt$ 
can be written in terms of an integral of $v$ also. Then
$\tau_{s0}=\int_{1}^{0} \frac{e^\nu}{\dot{v}}dv$. If we consider 
the part of the integral where $v<<1$, it is given by 
$\int_{0}^{\epsilon}v^2dv$, where $\epsilon<<1$. This goes to zero. 
So the proper time $\tau_{s0}$ is finite. Thus for the above generic 
form of $M$, a non-simultaneous singularity may occur.
Such a singularity may also be spacelike or timelike in this 
case.

\section{APPENDIX A: General form of $M$}

Since $M(0,v)= \frac{m_0}{v^3}$, $M$ can in general be written 
in the form,
\begin{equation}
 M(r,v)= \frac{m_0}{v^3}+f_2(r)g(r,v)
\end{equation}
where $g(o,v)\neq 0$ and $f_2(0)=0$. Then $M,_v=-\frac{3m_0}{v^4}
+f_2(r)g,_v $ and $M,_r= f_2'(r)g(r,v)+f_2(r)g,_r$. 
>From \eqref{eq:eos}, 
\begin{equation}
 v'= \frac{v^4(3f_2(r)g+rf_2'(r)g+rf_2(r)g,_r+vf_2(r)g,_v)}
{2r(3m_0-f_2(r)v^4g,_v)}
\end{equation}
This gives 
\be
R'=v+rv'=v[1+\frac{v^3(3f_2(r)g+rf_2'(r)g+rf_2(r)g,_r+vf_2(r)g,_v)}
{2(3m_0-f_2(r)v^4g,_v)}]
\ee
>From \eqref{Gequn}, we have 
\begin{eqnarray}
 G=\frac{2}{f^2(r)}[3m_0+v^3(3f_2(r)g+rf_2'(r)g+rf_2(r)g,_r+
vf_2(r)g,_v)+\nonumber \\
\frac{v^6}{4}\frac{(3f_2(r)g+rf_2'(r)g+rf_2(r)g,_r+vf_2(r)g,_v)^2}
{(3m_0-f_2(r)v^4g,_v)}] 
\label{apG}
\end{eqnarray}
For regular epochs (i.e. $v\neq0$),$M$ and its derivatives are bounded. 
In the integrand of $t_s(r)$, we have in the denominator, a factor 
$[\frac{v}{r^2}(G-1)+M]^{\frac{1}{2}}$, which must be finite when $r$ 
goes to zero. Since $M(0,v)$ is finite, $\frac{v}{r^2}(G-1)$ must 
be finite. This means $\lim{r \to 0} \frac{1}{r^2}(G-1)$ is finite. 
We can write the quantity $\frac{1}{r^2}(G-1)$ as,
\begin{eqnarray}
\frac{1}{r^2}(G-1)= 
(\frac{6m_0}{r^2f^2(r)}-\frac{1}{r^2})+\frac{2v^3}{r^2f^2(r)}
(3f_2(r)g+rf_2'(r)g+rf_2(r)g,_r+vf_2(r)g,_v)+\nonumber \\
\frac{v^6}{2r^2f^2(r)}
\frac{(3f_2(r)g+rf_2'(r)g+rf_2(r)g,_r+vf_2(r)g,_v)^2}{(3m_0-f_2(r)v^4g,_v)}
\end{eqnarray}
Now the first term in the bracket and the coefficients of $v^3$ 
and $v^6$ have to be zero separately. This gives,
\be
f^2(r)=\frac{6m_0}{1+r^2f_1(r)}
\ee
The coefficient of $v^3$ is $\frac{1}{6m_0}(f_1(r)+\frac{1}{r^2})
(3f_2(r)g+rf_2'(r)g+rf_2(r)g,_r+vf_2(r)g,_v)$.
This implies that $\lim{r \to 0}\frac{f_2(r)}{r^2}$ and 
$\lim{r \to 0}\frac{rf_2'(r)}{r^2}$ are bounded. This implies that as $r$ 
goes to zero, $f_2(r)$ goes to zero atleast as fast as $r^2$.

\section{APPENDIX B: Ruling out solutions for which 
$O[g_o(v)]>O(\frac{1}{v^7})$}

In this case, $\lim_{v \to 0}[\frac{v}{r^2}(G-1)+M]_{r=0}= 
\frac{v^4}{3m_0}g_0(v)$. Let, $\lim_{v \to 0} g_o(v)= \frac{c_3}{v^{7+\delta}}$, 
where $\delta>0$ and $c_3$ is some constant. Using this and 
\eqref{nu0v}, we get, $t_{sd}(0)=\int_0^\epsilon \frac{\sqrt{v}}
{v^3f_3(v)(\frac{c_3}{v^{3+\delta}})^{\frac{1}{2}}} dv$.  This can be 
written as $t_{sd}(0)=\frac{1}{f_3(0)}[v^{\frac{\delta}{2}}]_0^\epsilon$. 
This is vanishingly small. So $t_s(0)$ is finite.

However, for $r>0$, $\lim_{v \to 0} [\frac{v}{r^2}(G-1)+M]_{r=0}= 
\frac{r^2c_3}{v^{7+\delta}}$ and $\lim_{v \to 0} e^{\nu(r,v)}=v^{5+ 
\frac{\delta}{2}} f_3(r,v)$. 
Putting them together, we get, $t{sd}(r)= \frac{1}{r\sqrt{c_3}}\int_0^\epsilon 
\frac{1}{vf_3(r,v)}dv$.
This implies that $t_s(r)$ is divergent when $r>0$.

Now there are the following possibilities. Firstly, the singularity 
can be non-simultaneous, i.e, $v'>0$ at the epoch when some $r>0$ shell 
reaches the singularity. In this case, since $t_s(r)=\infty$ for $r>0$, 
arguing in the same way as in Prop.3; it can be shown that $\tau_s(r)$ 
will be divergent for some value of $r$. This means that the singularity 
curve $\tau_s(r)$ is discontinuous. So this possibility is ruled out.

There is also another possibility that all the shells $r>0$ hit 
the singularity at the same time, i.e. $v'=0$ for $r>0$ at that epoch. 
Since $v'=0$ at the epoch of singularity of some $r>0$ shell, all 
$r>0$ shells reach the singularity at the same epoch. The proper time 
taken along a shell with a very small value of $r$ would be close to 
the proper time taken along the $r=0$ shell to reach the singularity.
If the $\tau_s(r)$ is continuous, then it follows that at 
$\tau_{0}=\tau_{s0}$, all the shells reach the singularity. But 
this contradicts the fact that $t_s(0)<t_s(r)$ for $r>0$. So this 
second possibility is ruled out.

\end{document}